\documentclass[%
 reprint,
 amsmath,amssymb,
 aps,
 pre,
]{revtex4-2}

\usepackage{graphicx}
\usepackage{dcolumn}
\usepackage{bm}

\usepackage{braket}
\usepackage{stackrel}
\usepackage{subcaption}
\usepackage[hidelinks]{hyperref}

\begin{document}

\title{Escape Time with Quenched Extrinsic Fluctuations in Positive Feedback Systems}

\author{Manuel Eduardo Hernández-García}
\email{manuel.hernandezgarcia@viep.com.mx}
\author{Natalia López-Paleta}%
 \email{nlpaleta23@gmail.com}
 
\author{Jorge Velázquez-Castro}
\email{jorgevc@fcfm.buap.mx}
\affiliation{%
 Facultad de Ciencias Físico Matemáticas, Benemérita Universidad Autónoma de Puebla, Heroica Puebla de Zaragoza 72570, México. }%

\date{\today}

\begin{abstract}
Gene regulatory networks with positive feedback often exhibit multiple stable steady states representing distinct cellular phenotypes, allowing stochastic fluctuations to drive phenotypic switching between basins of attraction. Such transitions are commonly characterized by escape-time statistics. While previous studies have typically analyzed intrinsic fluctuation or dynamical environmental fluctuation separately, the combined effect of intrinsic biochemical fluctuation and quenched (slowly varying) extrinsic cell-to-cell parameter heterogeneity on population-level switching dynamics remains less understood.
In this study, we formulate a theoretical framework to compute escape-time moments by combining the backward chemical Fokker--Planck equation with population averaging over a stationary distribution of kinetic parameters. We apply this approach to two paradigmatic bistable motifs---positive autoregulation and the genetic Toggle Switch---and quantify how quenched extrinsic heterogeneity modifies switching times across a population of autonomous cells.
Our results show that extrinsic parameter variability can substantially alter the mean switching time and, more prominently, increase its dispersion, producing highly heterogeneous switching kinetics across genetically identical cells. These findings highlight the importance of accounting for cell-to-cell parameter heterogeneity when characterizing stochastic phenotypic switching in biological populations.
\end{abstract}

\maketitle


\section{Introduction}

Biological systems inherently exhibit fluctuations, and the concentrations of biochemical species vary continuously over time \cite{Paul, Schiavon}. These fluctuations are commonly classified as intrinsic or extrinsic. Intrinsic fluctuations arise from the stochastic nature of biochemical reactions and scale inversely with the system size \cite{Gar}, whereas extrinsic fluctuations are associated with variability in kinetic parameters \cite{Lei, Shahreza}, originating from environmental influences or cell-to-cell heterogeneity in cellular state \cite{extri, Eloe}. Because biological populations are neither isolated nor strictly homogeneous, both sources of variability are inevitably present.

Positive feedback motifs are commonly found in gene regulatory networks (GRNs) and play a fundamental role in processes such as cellular differentiation, decision-making, and adaptation \cite{Strasser, Alon, Pino, Yap}. These systems can exhibit bistability, characterized by the coexistence of two stable steady states representing distinct phenotypes. At the stochastic level, bistability gives rise to bimodal probability distributions whose modes correspond to these phenotypic states \cite{James, Perez}.

Fluctuations can induce spontaneous transitions between these modes, driving phenotypic switching from one basin of attraction to another. Such transitions are commonly characterized by the escape time, which quantifies the time required for the system to leave the basin of attraction of one stable state and reach another \cite{Poli, Kur, Gar, schnoerr2017efficient}. This phenomenon is of considerable interest because many biological processes rely on stochastic phenotypic switching, including cellular differentiation \cite{Yap} and antibiotic resistance \cite{Certain}. Likewise, certain pathological transitions, such as those observed in cancer, can be framed as switches between healthy and aberrant cellular states \cite{Well, Strasser}. More broadly, this framework also helps elucidate how fluctuations drive mutations and evolutionary dynamics \cite{Good}.

In many studies of escape times in GRNs, extrinsic fluctuations are neglected or not explicitly incorporated \cite{Erban, Wang}. In this work, our objective is to quantify how quenched (slowly varying) extrinsic fluctuations—representing cell-to-cell parameter heterogeneity across a population of autonomous cells—modify the statistics of phenotypic switching times. When parameter variations occur on timescales much longer than the switching process, each cell effectively evolves with fixed kinetic parameters drawn from a population distribution. We expect this extrinsic variability to substantially alter the population-level escape dynamics compared with the purely intrinsic case, both by introducing an additional source of dispersion \cite{extri, Eloe} and by shifting the locations of the stationary states and transition barriers across individuals \cite{Assaf}.

We analyze two representative GRNs that play important roles in a wide range of biological processes \cite{Schiavon, Jong}. The first is a positively autoregulated gene \cite{Assaf, Liu2004fluctuations} that exhibits two stable steady states. For this system, we compute the escape-time statistics associated with transitions between phenotypes under two scenarios: considering only intrinsic fluctuations, and incorporating both intrinsic fluctuations and quenched extrinsic parameter heterogeneity. The second system is a Toggle Switch \cite{Cantor}, a two-dimensional bistable model. Although this motif has been extensively studied, previous works have mainly focused on intrinsic fluctuations \cite{Barzel, Kim} or L\'evy noise \cite{Xu}. Here, we explicitly incorporate quenched extrinsic fluctuations to characterize the resulting population-level switching times and their heterogeneity.

To address this problem, we solve a hierarchy of differential equations derived from the backward chemical Fokker--Planck equation (FPE) \cite{grimaFP, Lopez} for fixed kinetic parameters, and subsequently average the resulting escape-time moments over the stationary parameter distribution describing the heterogeneous population. This approach is computationally more efficient than direct stochastic simulation algorithms used to sample the chemical master equation \cite{Erban, Scott}, and provides a direct alternative to minimum action path methods \cite{Xu2} (which rely on an effective potential landscape) and transition path theory \cite{Liu2}. Moreover, the FPE framework offers a transparent way to separate intrinsic and extrinsic contributions to switching-time variability. Because the FPE is a diffusion approximation of the underlying discrete stochastic process, its predictions become increasingly accurate in the large-system-size regime where intrinsic fluctuations are relatively small \cite{grimaFP}.

The remainder of this paper is organized as follows. In Section \ref{section2}, we review the forward and backward chemical FPE and introduce the quenched extrinsic parameter distribution describing cell-to-cell variability. In Section \ref{section3}, we formulate the differential equations and population-averaging framework used to compute the escape-time moments under both intrinsic and extrinsic fluctuations. Section \ref{Section4} is devoted to the analysis of a positively autoregulated system and its switching-time statistics. In Section \ref{Section5}, we study the Toggle Switch and quantify how extrinsic heterogeneity modifies its escape dynamics. Finally, our conclusions are presented in Section \ref{section6}.

\section{Backward Chemical Fokker--Planck Equation and Quenched Extrinsic Heterogeneity} \label{section2}

We first revisit the forward chemical FPE for a biochemical system with fixed kinetic parameters \cite{Gar,grimaFP}, and then incorporate quenched extrinsic variability across a heterogeneous cell population.

Consider a well-mixed biochemical system composed of $N$ chemical species $\mathcal{S}_l$ ($l \in \{1,2,\ldots, N\}$) interacting through $M$ reactions $\mathcal{R}_j$ ($j \in \{1,2,\ldots, M\}$):
\begin{align}
    \mathcal{R}_j : \sum_{l=1}^{N} \alpha_{jl} \mathcal{S}_l \stackbin[]{k_{j}}{\rightarrow} \sum_{l=1}^{N} \beta_{jl} \mathcal{S}_l, \nonumber
\end{align}
where $\alpha_{jl}$ and $\beta_{jl}$ are non-negative stoichiometric integers, $k_j$ are the kinetic parameters, and the stoichiometric matrix has entries $\Gamma_{lj} = \beta_{jl} - \alpha_{jl}$. According to the law of mass action, the macroscopic reaction rates are given by \cite{Gar}
\begin{align}
    R_j(\mathbf{s}, k_j) = k_j \prod_{l=1}^{N} s_l^{\alpha_{jl}}, \nonumber
\end{align}
where $\mathbf{s} = (s_1, s_2, \ldots, s_N)$ denotes the vector of species concentrations. Let $\Omega$ be the system size \cite{HillD}. In the large-$\Omega$ regime, the stochastic dynamics of the concentrations for a fixed parameter vector $\mathbf{k} = (k_1, k_2, \ldots, k_M)$ can be approximated by the forward chemical FPE \cite{Gar,grimaFP}:
{\small
\begin{widetext}
\begin{align}
    \frac{\partial P_{\mathbf{k}}(\mathbf{s}^{(f)},t^{(f)}|\mathbf{s}^{(i)},t^{(i)})}{\partial t^{(f)}} = & -\sum_{l_1=1}^{N} \frac{\partial}{\partial s^{(f)}_{l_1}} \left[ \sum_{j=1}^{M} \Gamma_{l_1j} R_j(\mathbf{s}^{(f)},k_j) P_{\mathbf{k}}(\mathbf{s}^{(f)},t^{(f)}|\mathbf{s}^{(i)}, t^{(i)}) \right] \nonumber \\
    & + \frac{1}{2} \sum_{l_1=1}^{N} \sum_{l_2=1}^{N} \frac{\partial^2}{\partial s^{(f)}_{l_1} \partial s^{(f)}_{l_2}} \left[ \sum_{j=1}^{M} \frac{\Gamma_{l_1j}\Gamma_{l_2j} R_j(\mathbf{s}^{(f)},k_j)}{\Omega} P_{\mathbf{k}}(\mathbf{s}^{(f)},t^{(f)}|\mathbf{s}^{(i)}, t^{(i)}) \right], \label{2}
\end{align}
\end{widetext}}
where $P_{\mathbf{k}}(\mathbf{s}^{(f)},t^{(f)} \mid \mathbf{s}^{(i)},t^{(i)})$ is the conditional probability density of finding the system in state $\mathbf{s}^{(f)}$ at time $t^{(f)}$ given initial state $\mathbf{s}^{(i)}$ at time $t^{(i)}$. Eq. \eqref{2} accounts solely for intrinsic biochemical fluctuations.

To incorporate extrinsic fluctuations, we note that across a cell population, kinetic parameters vary from cell to cell due to environmental influences or slowly varying cellular states (such as differences in transcriptional machinery, cell size, or epigenetic state) \cite{extri, Eloe, Vastola}. The statistical distribution of $\mathbf{k}$ across the population can be generated by an underlying stochastic process governed by a forward FPE:
\begin{widetext}
\begin{align}
    \frac{\partial p(\mathbf{k},t)}{\partial t} = -\sum_{j_1=1}^{M} \frac{\partial}{\partial k_{j_1}} \left[ f_{j_1}(\mathbf{k}) p(\mathbf{k},t) \right] + \frac{1}{2}\sum_{j_1=1}^{M} \sum_{j_2=1}^{M} \frac{\partial^2}{\partial k_{j_1} \partial k_{j_2}} \left[ F_{j_1 j_2}(\mathbf{k}) p(\mathbf{k},t) \right], \label{3}
\end{align}
\end{widetext}
where $f_{j_1}(\mathbf{k})$ and $F_{j_1 j_2}(\mathbf{k})$ represent the drift and diffusion terms of the parameter dynamics, respectively (see Appendix \ref{A} for explicit Langevin/FPE examples \cite{Gar, Vastola}).

In this work, we consider the \emph{quenched} (static-disorder) extrinsic fluctuation regime, in which the correlation time $\tau_k$ of parameter fluctuations is much longer than the characteristic switching timescale ($\tau_k \gg T_1$), and the population has reached its stationary parameter distribution $p_{\mathrm{ss}}(\mathbf{k})$ satisfying Eq.~\eqref{3} with $\partial_t p_{\mathrm{ss}}(\mathbf{k}) = 0$ \cite{biondo2023out, keizer2019extending}. In this regime, each autonomous cell evolves during the switching event with a fixed parameter realization $\mathbf{k}$ drawn from $p_{\mathrm{ss}}(\mathbf{k})$. By the law of total probability, the population-averaged transition probability density is therefore given by

{\small
\begin{align}
    \mathcal{P}(\mathbf{s}^{(f)}, t^{(f)} \mid \mathbf{s}^{(i)}, t^{(i)}) = \int d\mathbf{k}\, P_{\mathbf{k}}(\mathbf{s}^{(f)}, t^{(f)} \mid \mathbf{s}^{(i)}, t^{(i)})\, p_{\mathrm{ss}}(\mathbf{k}). \nonumber
\end{align}}
Because $\mathbf{k}$ remains constant along each cell's trajectory, we can first analyze the escape problem for a fixed $\mathbf{k}$ and subsequently average over $p_{\mathrm{ss}}(\mathbf{k})$ to obtain the population-level switching statistics.

For escape-time calculations, it is convenient to work with the backward chemical FPE associated with Eq.~\eqref{2}. Setting $t^{(i)}=0$, $t^{(f)}=t$, and initial condition $P_{\mathbf{k}}(\mathbf{s}^{(f)},0 \mid \mathbf{s}^{(i)},0) = \delta(\mathbf{s}^{(f)}-\mathbf{s}^{(i)})$, the time-homogeneous backward FPE reads \cite{Gar}
{\small
\begin{widetext}
\begin{align}
    \frac{\partial P_{\mathbf{k}}(\mathbf{s}^{(f)},t|\mathbf{s}^{(i)},0)}{\partial t} = & \sum_{l_1=1}^{N} \sum_{j=1}^{M} \Gamma_{l_1j} R_j(\mathbf{s}^{(i)},k_j) \frac{\partial P_{\mathbf{k}}(\mathbf{s}^{(f)},t|\mathbf{s}^{(i)},0)}{\partial s^{(i)}_{l_1}} \nonumber \\
    & + \frac{1}{2} \sum_{l_1=1}^{N} \sum_{l_2=1}^{N} \sum_{j=1}^{M} \frac{\Gamma_{l_1j}\Gamma_{l_2j} R_j(\mathbf{s}^{(i)},k_j)}{\Omega} \frac{\partial^2 P_{\mathbf{k}}(\mathbf{s}^{(f)},t|\mathbf{s}^{(i)},0)}{\partial s^{(i)}_{l_1} \partial s^{(i)}_{l_2}}, \label{FPE_ext}
\end{align}
\end{widetext}}
which describes the evolution of the conditional probability with respect to the initial concentration state $\mathbf{s}^{(i)}$ for a given cell with parameters $\mathbf{k}$.

\section{Escape Time with Quenched Extrinsic Fluctuations} \label{section3}

\begin{figure}[h!t]
\centering
\includegraphics[width=0.85\linewidth]{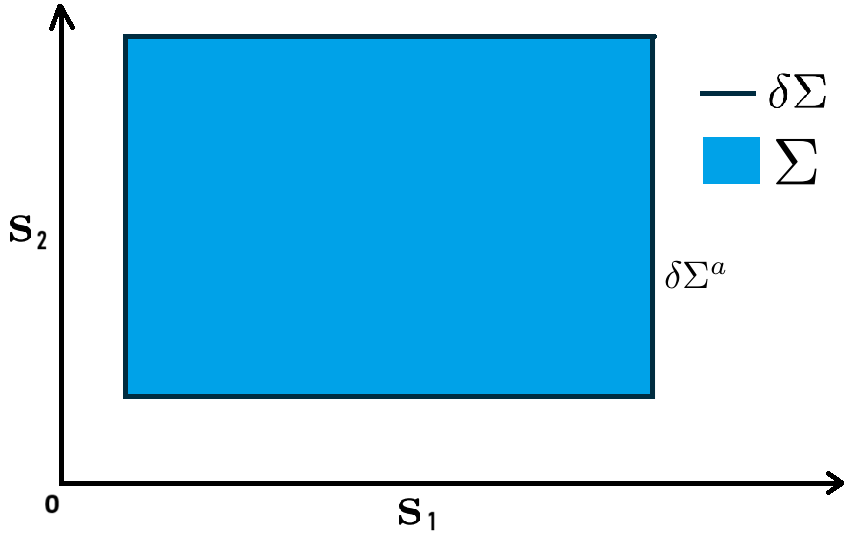}
\caption{\textbf{Escape region and its boundaries.} Schematic of the escape region $\Sigma$ delimited by its boundary $\delta \Sigma$. An absorbing barrier $\delta \Sigma^a$ marks the threshold for leaving $\Sigma$, while $\delta \Sigma^r$ denotes reflecting boundaries (e.g., at zero concentration).} \label{fig1}
\end{figure}

We now compute the switching-time statistics for a population of autonomous cells transitioning out of a basin of attraction $\Sigma$ with boundary $\delta\Sigma$ (Fig.~\ref{fig1}) under both intrinsic fluctuation and quenched extrinsic parameter heterogeneity \cite{Poli, Kur, Gar, biondo2023out}.

For a single cell with fixed kinetic parameters $\mathbf{k}$ starting at $\mathbf{s}^{(i)} \in \Sigma$ at $t=0$, the survival probability---that is, the probability that the cell has not yet left $\Sigma$ by time $t$---is
\begin{align}
    G(\mathbf{s}^{(i)}, \mathbf{k}, t) = \int_{\Sigma} d\mathbf{s}^{(f)}\, P_{\mathbf{k}}(\mathbf{s}^{(f)}, t \mid \mathbf{s}^{(i)}, 0). \label{G}
\end{align}
Integrating the backward FPE \eqref{FPE_ext} over $\mathbf{s}^{(f)} \in \Sigma$, we find that $G(\mathbf{s}^{(i)}, \mathbf{k}, t)$ obeys
\begin{widetext}
\begin{align}
    \frac{\partial G(\mathbf{s}^{(i)}, \mathbf{k}, t)}{\partial t} = \sum_{l_1=1}^{N} \sum_{j=1}^{M} \Gamma_{l_1j} R_j(\mathbf{s}^{(i)}, k_j) \frac{\partial G(\mathbf{s}^{(i)}, \mathbf{k}, t)}{\partial s^{(i)}_{l_1}} + \frac{1}{2} \sum_{l_1=1}^{N} \sum_{l_2=1}^{N} \sum_{j=1}^{M} \frac{\Gamma_{l_1j}\Gamma_{l_2j} R_j(\mathbf{s}^{(i)}, k_j)}{\Omega} \frac{\partial^2 G(\mathbf{s}^{(i)}, \mathbf{k}, t)}{\partial s^{(i)}_{l_1} \partial s^{(i)}_{l_2}}, \label{6}
\end{align}
\end{widetext}
subject to the initial condition and absorbing boundary condition:

\begin{subequations}
\begin{align}
    G(\mathbf{s}^{(i)}, \mathbf{k}, 0) &=
    \begin{cases}
        1 & \text{if } \mathbf{s}^{(i)} \in \Sigma, \\
        0 & \text{elsewhere},
    \end{cases} \label{G-Border0} \\
    G(\mathbf{s}^{(i)}, \mathbf{k}, t) &= 0 \quad \text{if } \mathbf{s}^{(i)} \in \delta\Sigma^a. \label{G-Border}
\end{align}
\end{subequations}

Across a heterogeneous population with quenched parameter distribution $p_{\mathrm{ss}}(\mathbf{k})$, the fraction of cells that remain in $\Sigma$ at time $t$ is the population-averaged survival probability
\begin{align}
    \bar{G}(\mathbf{s}^{(i)}, t) = \int d\mathbf{k}\, p_{\mathrm{ss}}(\mathbf{k})\, G(\mathbf{s}^{(i)}, \mathbf{k}, t). \nonumber
\end{align}
Since the cumulative probability of having escaped by time $t$ is $1 - \bar{G}(\mathbf{s}^{(i)}, t)$, the probability density of switching times across the population is $-\partial_t \bar{G}(\mathbf{s}^{(i)}, t)$. Consequently, the $\rho$-th moment of the population escape time is given exactly by
\begin{align}
    T_{\rho}(\mathbf{s}^{(i)}) = -\int_{0}^{\infty} t^{\rho}\, \partial_t \bar{G}(\mathbf{s}^{(i)}, t)\, dt. \label{Time}
\end{align}
Interchanging the order of integration and integrating by parts (using $G(\mathbf{s}^{(i)}, \mathbf{k}, 0) = 1$ and assuming $t^{\rho} G(\mathbf{s}^{(i)}, \mathbf{k}, t) \to 0$ as $t \to \infty$), we obtain
\begin{subequations} \label{11}
\begin{align}
    T_{\rho}(\mathbf{s}^{(i)}) &= \int d\mathbf{k}\, p_{\mathrm{ss}}(\mathbf{k})\, T_{\rho}(\mathbf{s}^{(i)}, \mathbf{k}), \\
    T_{\rho}(\mathbf{s}^{(i)}, \mathbf{k}) &= -\int_{0}^{\infty} t^{\rho}\, \partial_t G(\mathbf{s}^{(i)}, \mathbf{k}, t)\, dt \nonumber \\
    &= \rho \int_{0}^{\infty} t^{\rho-1} G(\mathbf{s}^{(i)}, \mathbf{k}, t)\, dt,
\end{align}
\end{subequations}
with $T_0(\mathbf{s}^{(i)}, \mathbf{k}) = 1$. Multiplying Eq.~\eqref{6} by $\rho t^{\rho-1}$ and integrating over $t \in [0, \infty)$ yields the standard recursive hierarchy of partial differential equations for the conditional escape-time moments $T_{\rho}(\mathbf{s}^{(i)}, \mathbf{k})$ \cite{Gar}:
\begin{widetext}
\begin{align}
    -\rho T_{\rho-1}(\mathbf{s}^{(i)}, \mathbf{k}) = \sum_{l_1=1}^{N} \sum_{j=1}^{M} \Gamma_{l_1j} R_j(\mathbf{s}^{(i)}, k_j) \frac{\partial T_{\rho}(\mathbf{s}^{(i)}, \mathbf{k})}{\partial s_{l_1}^{(i)}} + \frac{1}{2} \sum_{l_1=1}^{N} \sum_{l_2=1}^{N} \sum_{j=1}^{M} \frac{\Gamma_{l_1j}\Gamma_{l_2j} R_j(\mathbf{s}^{(i)}, k_j)}{\Omega} \frac{\partial^2 T_{\rho}(\mathbf{s}^{(i)}, \mathbf{k})}{\partial s_{l_1}^{(i)} \partial s_{l_2}^{(i)}}. \label{switch-time2}
\end{align}
\end{widetext}
To solve Eq.~\eqref{switch-time2} numerically, we impose absorbing conditions at $\delta\Sigma^a(\mathbf{k})$ and reflecting conditions at $\delta\Sigma^r$ (such as zero-concentration boundaries, since concentrations cannot become negative). Integrating the boundary conditions for $G(\mathbf{s}^{(i)}, \mathbf{k}, t)$ over time yields
\begin{widetext}
\begin{subequations} \label{13}
\begin{align}
    \sum_{l_1=1}^{N} \sum_{l_2=1}^{N} n_{l_1} \left( \sum_{j=1}^{M} \frac{\Gamma_{l_1j}\Gamma_{l_2j} R_j(\mathbf{s}^{(i)}, k_j)}{\Omega} \right) \frac{\partial T_{\rho}(\mathbf{s}^{(i)}, \mathbf{k})}{\partial s_{l_2}^{(i)}} &= 0 \quad \text{if } \mathbf{s}^{(i)} \in \delta\Sigma^r, \\
    T_{\rho}(\mathbf{s}^{(i)}, \mathbf{k}) &= 0 \quad \text{if } \mathbf{s}^{(i)} \in \delta\Sigma^a(\mathbf{k}),
\end{align}
\end{subequations}
\end{widetext}
where $n_{l_1}$ denotes the components of the unit vector normal to $\delta\Sigma^r$. Note that the absorbing boundary $\delta\Sigma^a(\mathbf{k})$---representing the separatrix between basins of attraction---generally depends on $\mathbf{k}$.

In this study, we focus on the mean escape time $T_1(\mathbf{s}^{(i)})$ and the second moment $T_2(\mathbf{s}^{(i)})$, from which the population variance $\sigma_T^2(\mathbf{s}^{(i)}) = T_2(\mathbf{s}^{(i)}) - [T_1(\mathbf{s}^{(i)})]^2$ and standard deviation $\sigma_T(\mathbf{s}^{(i)})$ are obtained. In the quenched regime, the law of total variance provides a transparent decomposition:
\begin{align}
    \sigma_T^2(\mathbf{s}^{(i)}) = \mathbb{E}_{\mathbf{k}}\!\left[\sigma_T^2(\mathbf{s}^{(i)}, \mathbf{k})\right] + \mathrm{Var}_{\mathbf{k}}\!\left[T_1(\mathbf{s}^{(i)}, \mathbf{k})\right], \label{total_var}
\end{align}
where the first term is the population average of the intrinsic escape-time variance and the second term captures the additional dispersion induced by cell-to-cell variability in the mean switching time.

Finally, in bistable regulatory networks, a parameter distribution $p_{\mathrm{ss}}(\mathbf{k})$ may assign nonzero probability mass to monostable regions of parameter space where bistable switching does not occur. Denoting the bistable region by $\mathcal{B}$, the fraction of switching-competent (bistable) cells is $f_{\mathrm{bi}} = \int_{\mathcal{B}} p_{\mathrm{ss}}(\mathbf{k})\, d\mathbf{k}$. In such cases, we condition Eq.~\eqref{11} on the bistable subpopulation by replacing $p_{\mathrm{ss}}(\mathbf{k})$ with the normalized conditional density $\tilde{p}_{\mathrm{ss}}(\mathbf{k}) = p_{\mathrm{ss}}(\mathbf{k}) / f_{\mathrm{bi}}$ for $\mathbf{k} \in \mathcal{B}$ (and $0$ otherwise), while noting that the remaining fraction $1 - f_{\mathrm{bi}}$ exhibits monostable dynamics.

\section{Model I: Positive Auto-regulation} \label{Section4}

\begin{figure} [h!t]
\centering
\includegraphics[width=0.45\linewidth]{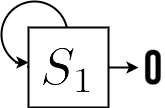}
\caption{ \textbf{Positive Auto-regulation.} In this model the molecule $S_1$ auto-activates its production. $S_1$ is also degraded. The loop represents a system with positive feedback. }
\label{fig2}
\end{figure}

Positive autoregulation is a model of gene expression, in which the rate of expression is controlled by the gene itself \cite{Liu2004fluctuations}, as illustrated in Fig. \ref{fig2}. This network motif plays a key role in GRNs and contributes to system stability and robustness \cite{Pino}. Research has shown that positive autoregulation is critical in various biological systems. For instance, in yeast, the PDR3 gene, which is associated with drug resistance, employs positive autoregulation through specific promoter binding sites \cite{Dela}.   
 
This model, representing the simplest positive-feedback system, provides a natural starting point for studying phenotypic switching times under intrinsic and quenched extrinsic fluctuations. Understanding its dynamics establishes a foundation for investigating more complex regulatory networks involving additional molecular species while retaining positive-feedback interactions.

The reactions in this model are as follows,
\begin{align}
    & 0  \stackbin{k_1^* }{\rightarrow} S_1, & S_1  \stackbin{k_3}{\rightarrow} 0, \nonumber
\end{align}
the first reaction is auto-activation via positive regulation, and the other is degradation. The reaction rates are: 
\begin{align}
    R_1 =&  k_1^*=  \frac{k_1 s_1^2 +k_2 }{1+s_1^2}, &  R_2 =& k_3 s_1, \nonumber
\end{align}
where $s_1$ denotes the concentration of $S_1$, $k_1$ represents the maximum regulation-induced synthesis rate, $k_2$ is the basal synthesis rate, and $k_3$ is the degradation rate. Here, we employ a Hill function with a dissociation constant equal to one and a Hill coefficient equal to two \cite{Manu2026}. The corresponding stoichiometric matrix is given by
\begin{align}
 \Gamma_{ij}&= \begin{pmatrix}
		1& -1 
	\end{pmatrix}. \nonumber
\end{align}

\begin{figure}
    \centering
 \includegraphics[width=0.85\linewidth]{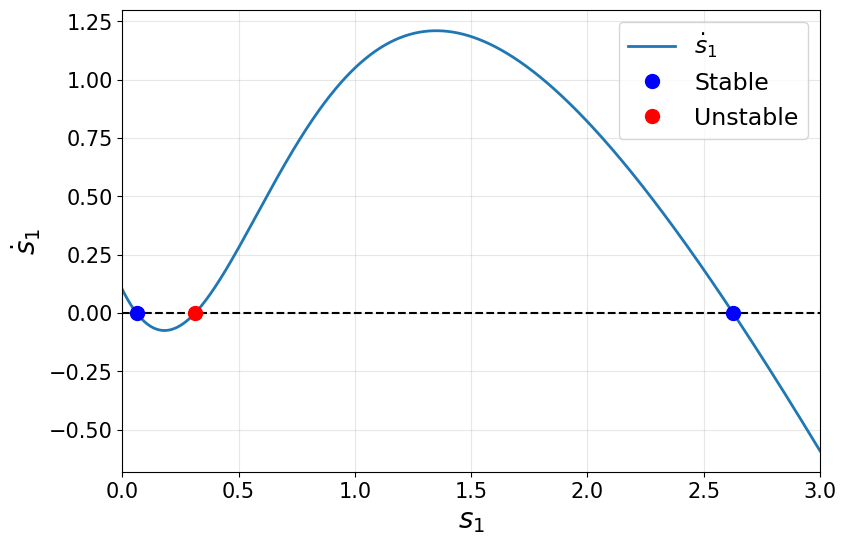}
    \caption{\textbf{Phase Space of Positive Auto-regulation.} In this figure, we plot the phase space of the positive autoregulation model. We observe that the system exhibits bistability, characterized by two stable stationary states (blue) separated by one unstable stationary state (red). We used $k_1=6$, $k_2=0.1$ and $k_3=2$. }
    \label{fig3}
\end{figure}

This system may be represented by the deterministic model
\begin{align}
    \frac{\partial s_1}{\partial t} =&\frac{k_1 s_1^2 +k_2 }{1+s_1^2} -  k_3 s_1 , \nonumber
\end{align}
from this ordinary differential equation (ODE), we choose $k_1=6$, $k_2=0.1$, and $k_3=2$ to obtain a bistable system with stationary states at (see Fig. \ref{fig3})
\begin{subequations}
\begin{align}
    \bar{s_1}=&0.061,& \bar{s_2}=&2.626 & \text{stable},  \label{14} \\
    \bar{s_3}=& 0.313,&&& \text{unstable}.
\end{align}
\end{subequations}

Biologically, the state $\bar{s_1}$ may represent a transcriptionally inactive phenotype, whereas the state $\bar{s_2}$ represents an active phenotype. Therefore, the escape process describes stochastic phenotypic activation driven by intrinsic and quenched extrinsic fluctuations. 

In subsections IV.A and IV.B, we calculate the escape time under only intrinsic fluctuations, and when both intrinsic fluctuations and quenched extrinsic heterogeneity are considered. Eq. \eqref{switch-time2} is solved numerically using the finite difference method, subject to the boundary conditions given in Eq. \eqref{13}.

\subsection{Intrinsic fluctuations}

\begin{figure*}[h!t]
\centering
\includegraphics[width=0.435\textwidth]{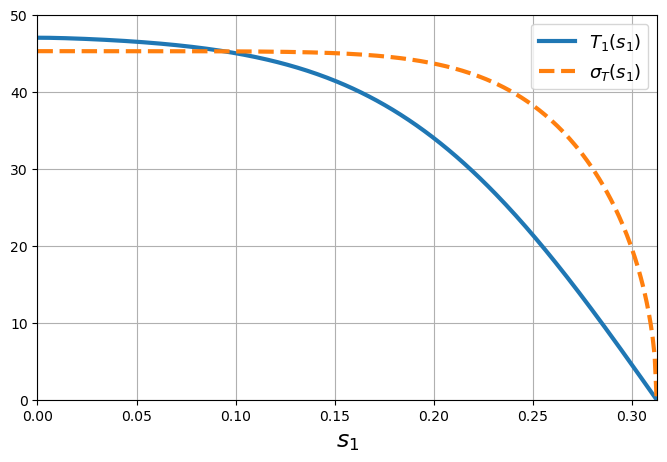}
\includegraphics[width=0.455\textwidth]{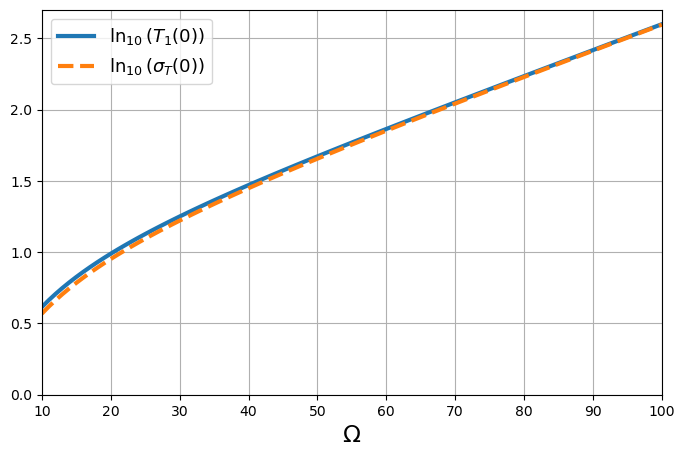}
\caption{ \textbf{Escape time for positive autoregulation under intrinsic fluctuations.} In the left panel, we present the mean escape time $T_1(s_1)$ and its standard deviation $\sigma_T(s_1)$ for $\Omega=50$. Both quantities have comparable magnitudes ($\sigma_T \approx T_1$), characteristic of nearly exponential escape-time distributions. In the right panel, we show the mean escape time and its standard deviation (on a $\log_{10}$ scale) for an initial concentration equal to zero and different values of $\Omega$. As $\Omega$ increases, both quantities grow exponentially while remaining of similar magnitude. We used $k_1=6$, $k_2=0.1$, and $k_3=2$.}
\label{fig4}
\end{figure*}

First, we consider only intrinsic fluctuations. Using the reaction rates and stoichiometric matrix, we obtain the differential equation for the $\rho$-moment escape time based on Eq. \eqref{switch-time2}, 

\begin{align}\label{burst_switch_time}
    -\rho T_{\rho-1}(s_1)=&  \left( \frac{k_1 s_1^2 +k_2 }{1+s_1^2} -  k_3 s_1 \right) \frac{\partial T_{\rho}( s_1)}{ \partial s_1} \nonumber \\
    &+  \frac{\left( \frac{k_1 s_1^2 +k_2 }{1+s_1^2} +  k_3 s_1 \right) }{2 \Omega}  \frac{\partial^2 T_{\rho}(s_1)}{ \partial s_1^2 }.
\end{align}
To solve this differential equation numerically, appropriate boundary conditions must be specified. A reflecting boundary arises because the concentration $s_1$ cannot take negative values, while an absorbing boundary is imposed when the system reaches the unstable stationary state $\bar{s}_3$ (see Fig. \ref{fig3}). The corresponding conditions are given by

\begin{subequations}\label{conditions_1}
\begin{align}
    \frac{\partial T_{\rho}(s_1)}{\partial s_1} =& 0 \quad \text{if } s_1=0 , \\
    T_{\rho}(s_1) =& 0  \quad \text{if } s_1=\bar{s}_3 .
\end{align} 
\end{subequations}

In Fig. \ref{fig4}, we present the mean escape time, $T_1(s_1)$, together with its standard deviation, $\sigma_T(s_1)$, for different initial conditions below the unstable stationary state $\bar{s}_3$. Both quantities decrease as the initial condition approaches $\bar{s}_3$ because the system requires less time to cross the barrier separating the two stable regions. Notably, the mean escape time and its standard deviation have comparable magnitudes ($\sigma_T \approx T_1$), which is characteristic of nearly exponential escape-time distributions associated with Kramers-like barrier crossing \cite{hanggi1990reaction}.

We now focus on the escape process from the zero-concentration state for different values of $\Omega$. The results are shown in the right panel of Fig. \ref{fig4}, where we plot the mean escape time together with its standard deviation on a base-10 logarithmic scale. As expected, both quantities increase exponentially with $\Omega$, reflecting the reduction in intrinsic fluctuations in larger systems, while maintaining $\sigma_T \approx T_1$. Our analysis begins at $\Omega=10$, because the FPE provides a reliable approximation only for sufficiently large values of $\Omega$.

\subsection{Intrinsic and quenched extrinsic fluctuations}

We now consider a heterogeneous cell population under the combined influence of intrinsic fluctuations and quenched extrinsic fluctuations. In particular, cell-to-cell variability is introduced through static heterogeneity in the kinetic parameters \cite{Shahreza, extri}. Here, we focus on variability in $k_1$, which is associated with the transcription process and is naturally subject to cell-to-cell fluctuations \cite{Vastola, Shahreza}. Additionally, variability in the degradation rate $k_3$ can also be considered, as discussed in \cite{Wang}.

To describe the population distribution of $k_1$, we take the stationary distribution of a slow Cox--Ingersoll--Ross process \cite{Vastola, Assaf}, which is a gamma distribution (see Appendix \ref{A}):

\begin{align}
    p_{\mathrm{ss}}(k_1) =& \frac{\left( \frac{a}{k_{0,1}} \right)^a}{\Gamma(a)} k_1^{a-1} e^{-\frac{a}{k_{0,1}} k_1},  \label{15}
\end{align}
where $k_{0,1}$ denotes the mean value of the kinetic parameter across the population, $\Gamma(a)$ is the Gamma function, and $a$ controls the magnitude of the extrinsic heterogeneity (smaller values of $a$ correspond to broader cell-to-cell variability). To facilitate a direct comparison with the purely intrinsic case, we set $k_{0,1}=6$. This distribution has mean $k_{0,1}$ and variance $k_{0,1}^{2}/a$, and is defined strictly on non-negative values of $k_1$, which is physically consistent for kinetic rates.

\begin{figure*}[h!t]
\centering
\includegraphics[width=0.32\textwidth]{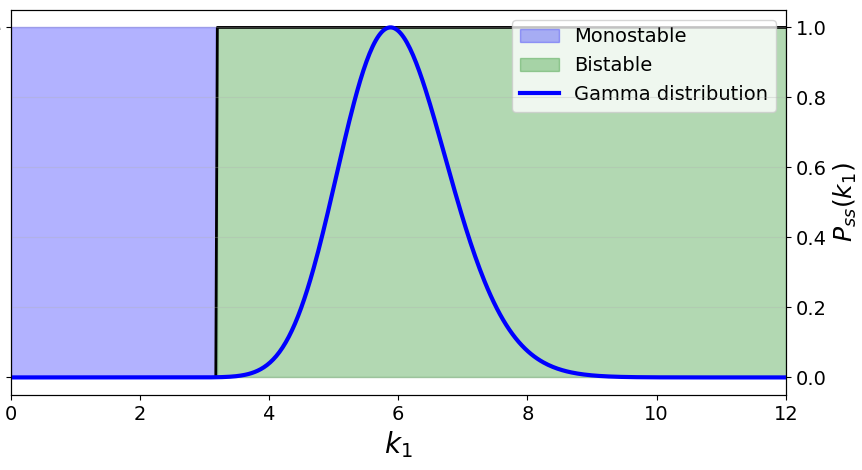}
\includegraphics[width=0.32\textwidth]{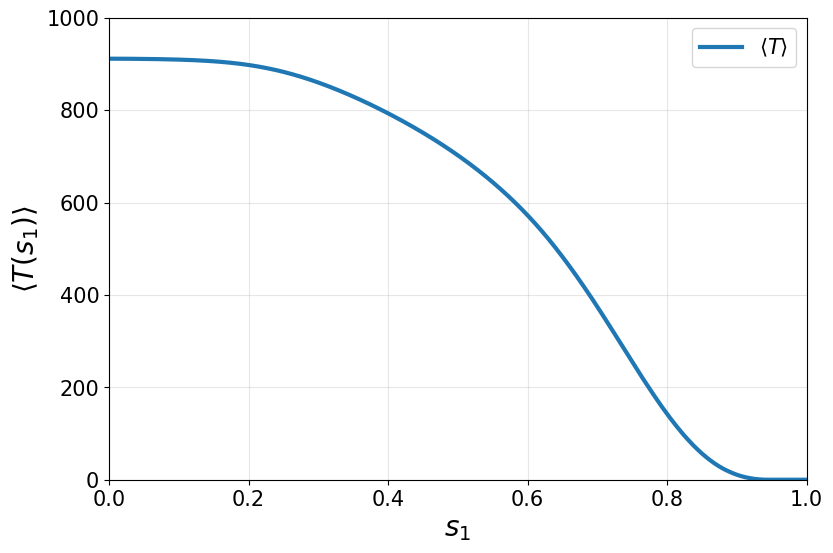}
\includegraphics[width=0.32\textwidth]{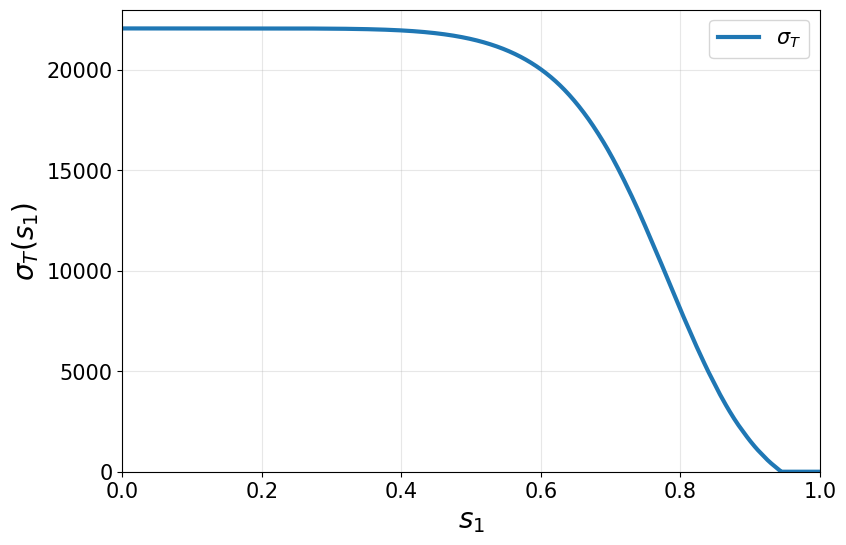}
\caption{ \textbf{Escape time for positive autoregulation under intrinsic and quenched extrinsic fluctuations.} In the left panel, we plot the probability distribution of $k_1$ together with the parameter regions in which the system is monostable or bistable. Although the distribution extends slightly into the monostable regime, we condition our analysis on the bistable subpopulation $\mathcal{B}$. In the other two panels, we present the population mean escape time and its standard deviation. Both quantities increase compared with the purely intrinsic case, with a particularly pronounced growth in the standard deviation ($\sigma_T \gg T_1$), indicating highly heterogeneous switching times across cells. We used $a=50$, $\Omega=50$, $k_{0,1}=6$, $k_2=0.1$, and $k_3=2$.}
\label{fig5}
\end{figure*}

\begin{figure*}[h!t]
\includegraphics[width=0.495\textwidth]{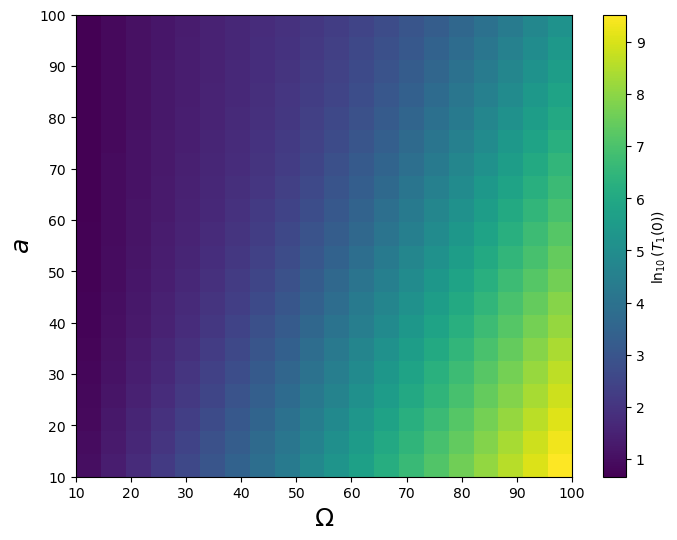}\hfill
\includegraphics[width=0.495\textwidth]{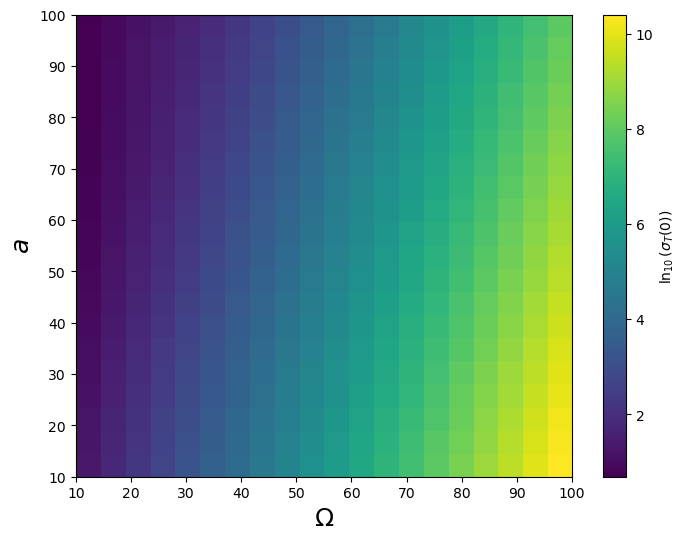}
\caption{\textbf{Escape time for positive autoregulation under different intensities of intrinsic and quenched extrinsic fluctuations.} 
In these panels, we compute the mean escape time and its standard deviation (on a $\log_{10}$ scale) for different values of $\Omega$ and $a$, using an initial concentration of zero. The mean escape time increases as $\Omega$ increases and decreases as $a$ increases. Larger values of $a$ reduce cell-to-cell variability, concentrating $p_{\mathrm{ss}}(k_1)$ around its mean $k_{0,1}$ and reducing the weight of parameter realizations associated with longer escape times. We used $k_{0,1}=6$, $k_2=0.1$, and $k_3=2$.}
\label{fig6}
\end{figure*}

In Fig. \ref{fig5} (left panel), we identify the monostable and bistable regimes as a function of $k_1$ and overlay the population distribution $p_{\mathrm{ss}}(k_1)$. For small values of $k_1$ below the saddle-node bifurcation, the active state disappears and cells remain monostable in the inactive phenotype (unable to switch). Following Section \ref{section3}, we condition Eq.~\eqref{11} on the switching-competent bistable subpopulation $\mathcal{B}$ using the normalized conditional distribution $\tilde{p}_{\mathrm{ss}}(k_1)$.

We solved Eq. \eqref{burst_switch_time} for each value of $k_1 \in \mathcal{B}$—with its corresponding $k_1$-dependent unstable threshold $\bar{s}_3(k_1)$, setting $T_{\rho}(s_1, k_1)=0$ for $s_1 \ge \bar{s}_3(k_1)$—and integrated the resulting moments over $k_1$ according to Eq. \eqref{11}. Note that in Fig. \ref{fig5}, $s_1$ extends up to $s_1 \approx 1$ because $\bar{s}_3(k_1)$ shifts upward as $k_1$ decreases toward the bifurcation boundary. Both the mean escape time and its standard deviation increase substantially compared with the purely intrinsic case (Fig. \ref{fig4}). In particular, the standard deviation exceeds the mean by more than an order of magnitude ($\sigma_T \gg T_1$).

This pronounced growth in dispersion is directly explained by the law of total variance in Eq.~\eqref{total_var}. Because each realization of $k_1$ shifts the location of the unstable state $\bar{s}_3(k_1)$ and the height of the effective barrier \cite{Assaf}, the conditional mean switching time $T_1(s_1, k_1)$ varies exponentially across cells. Consequently, the extrinsic variance term $\mathrm{Var}_{k_1}[T_1(s_1, k_1)]$ dominates the total variance, generating highly heterogeneous switching times across the population. Furthermore, conditioning on $\mathcal{B}$ excludes low-$k_1$ monostable cells while retaining high-barrier realizations that contribute heavily to both $T_1$ and $\sigma_T$.

To further investigate the interplay between intrinsic fluctuation and quenched extrinsic heterogeneity, we computed $T_1(0)$ and $\sigma_T(0)$ across values of $\Omega$ and $a$ (Fig. \ref{fig6}). Both quantities increase with $\Omega$ and decrease as $a$ increases. For small $a$ (strong cell-to-cell heterogeneity), $\sigma_T(0)$ becomes very large, signaling broadly dispersed switching kinetics across cells. As $a$ increases, the parameter distribution narrows around $k_{0,1}$, and the switching-time statistics gradually converge to the purely intrinsic limit.

\section{Model II: Toggle Switch} \label{Section5}
\begin{figure} [h!t]
\centering
\includegraphics[width=0.5\linewidth]{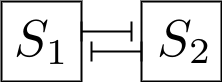}
\caption{ \textbf{Toggle Switch.} It is a model with two molecules, and each one represses the other, in such a way that it is a system with positive feedback. Additionally, each molecule is degraded.}
\label{toggles}
\end{figure}

A \textit{Toggle Switch} is a bistable system where two proteins mutually repress the synthesis of each other \cite{Cantor}, as illustrated in Fig. \ref{toggles}. Owing to its bistable nature, it can model a wide variety of cellular processes, either as an effective representation of an entire regulatory system or as a motif embedded within a larger regulatory network \cite{Alon, Jong}. For instance, in the context of antibiotic resistance, a \textit{Toggle Switch} may describe a subpopulation of bacteria in a state of susceptible or non-susceptible to antibiotics \cite{Certain}. Additionally, this motif has been used to describe cell fate decision processes \cite{Strasser}, and there is significant interest in understanding under which conditions these transitions occur. 

The reactions in this model are as follows,
\begin{align}
    & 0  \stackbin{k_1 H(s_2)}{\rightarrow} S_1, & S_1  \stackbin{k_2}{\rightarrow} 0, \nonumber \\
    & 0  \stackbin{k_3 H(s_1)}{\rightarrow} S_2, & S_2  \stackbin{k_4}{\rightarrow} 0, \nonumber
\end{align}
the upper reactions denotes the repression of $S_1$ by $S_2$, and the degradation of $S_1$. Similarly, for the other reactions. Where $H(s_i)$ is the Hill function
\begin{align}
    H(s_i)= \frac{1}{1+ s_i^2} \nonumber,
\end{align} 
$s_1$ and $s_2$ denote the concentration of $S_1$ and $S_2$, respectively. This Hill function has a Hill coefficient equal to two and a dissociation constant equal to one. Then, the reaction rates of the system are
\begin{align}
   R_1&= k_1 H(s_2) , &
    R_2= k_2 s_1, \nonumber \\
   R_3&= k_3 H(s_1) , &
    R_4= k_4 s_2,  \nonumber
\end{align}
and its corresponding stoichiometric matrix is
{\small
\begin{align}
 \Gamma_{ij}&= \begin{pmatrix}
		1& -1& 0 & 0  \\
		0& 0 & 1 & -1 
	\end{pmatrix}. \nonumber
\end{align}}

This system may be represented by the deterministic model
\begin{subequations} \label{18}
\begin{align}
    \frac{\partial s_1}{\partial t} =& k_1 H(s_2) - k_2 s_1 ,  \\
    \frac{\partial s_2}{\partial t} =& k_3 H(s_1) - k_4 s_2 , 
\end{align}
\end{subequations}

\begin{figure} [h!t]
\centering
\includegraphics[width=0.8\linewidth]{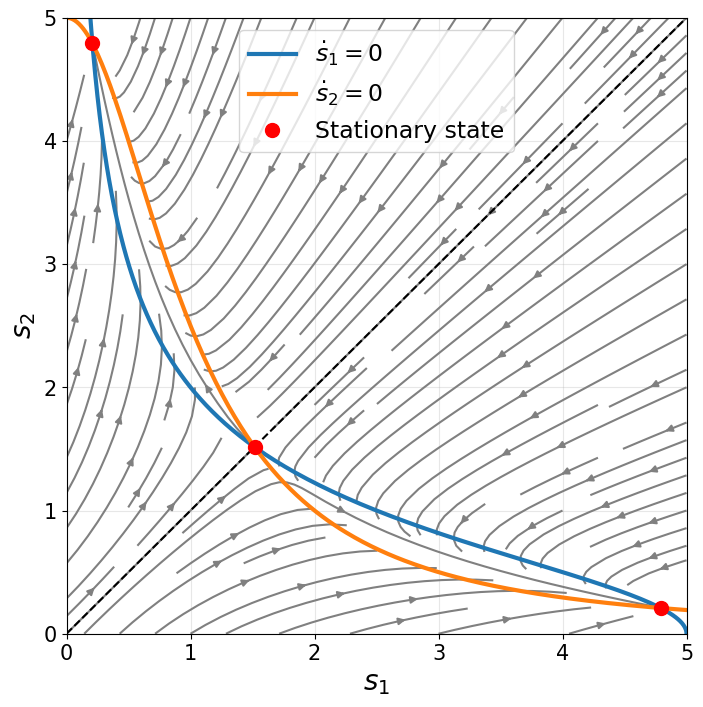}
\caption{ \textbf{Toggle switch vector field.} In this figure, we plot the vector field of the Toggle Switch. The red dots represent the stationary states and the black line represents the absorbing barrier $\delta\Sigma^a$. We used $k_1=k_3=2$ and $k_2=k_4=0.4$. }
\label{fig:VecF_toggle}
\end{figure}

We chose $k_1=k_3=2$, and $k_2=k_4=0.4$ to obtain a bistable system with stationary states at 
\begin{subequations}
\begin{align}
    \bar{s_1}=&(0.209,4.791),& \bar{s_2}=&(4.791,0.209) & \text{stable},   \\
    \bar{s_3}=& (1.516,1.516),&&& \text{unstable}.
\end{align}
\end{subequations}
In Fig. \ref{fig:VecF_toggle}, we can see the stationary states as red dots and the absorbing boundary as a black line. For symmetric parameters, the absorbing boundary ($s_1=s_2$) represents the separatrix beyond which the system deterministically relaxes toward the alternative phenotypic state \cite{Xu}.

\begin{figure*}
    \centering
    \includegraphics[width=0.325\linewidth]{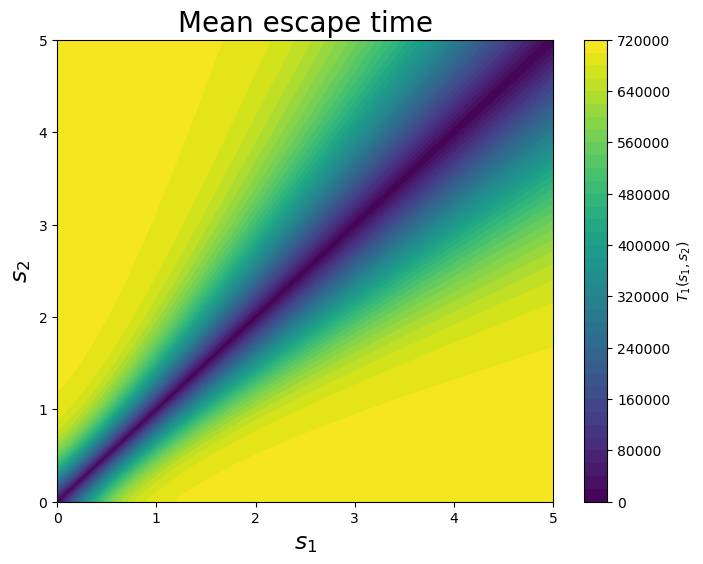}
    \includegraphics[width=0.325\linewidth]{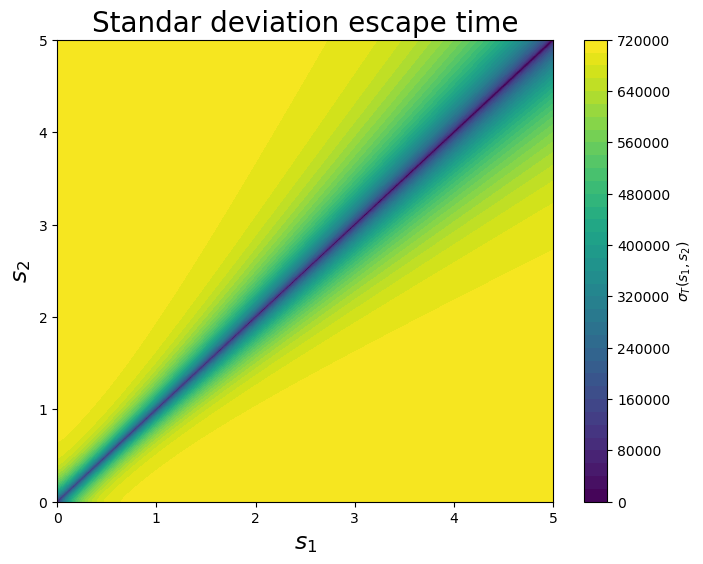}
    \includegraphics[width=0.325\linewidth]{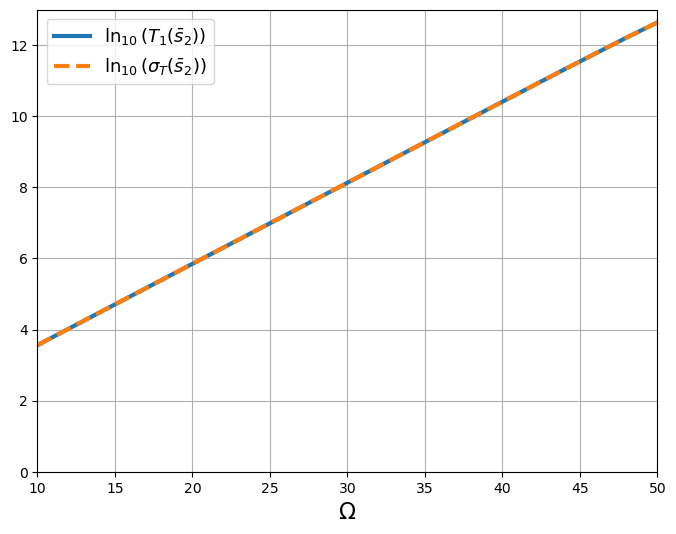}
    \caption{\textbf{Escape time for the toggle switch with intrinsic fluctuations.} In the first two panels, we plot the mean escape time $T_1(s_1,s_2)$ together with its standard deviation $\sigma_T(s_1,s_2)$ for $\Omega=20$. The mean escape time increases deeper inside each basin of attraction, and the standard deviation exhibits a spatial structure and magnitude comparable to the mean ($\sigma_T \approx T_1$). In the last panel, we plot $T_1(\bar{s}_2)$ and $\sigma_T(\bar{s}_2)$ on a $\log_{10}$ scale for trajectories starting from the stable stationary state $\bar{s}_2$. Both quantities are of comparable magnitude and grow exponentially with the system size $\Omega$ (appearing linear on the logarithmic scale). The parameter values used were $k_1=k_3=2$ and $k_2=k_4=0.4$.}
    \label{fig9}
\end{figure*}

\subsection{Intrinsic fluctuations}

To determine the escape time of this system with only intrinsic fluctuations, we derive the equation for the escape time by substituting the corresponding $\Gamma_{ij}$ and $R_i$ in Eq. \eqref{switch-time2}. Then we get,

\begin{widetext}
\begin{align}\label{toggle_switch_time}
    -{\rho} T_{\rho-1}(s_1, s_2) =& ( k_1 H(s_2) - k_2 s_1) \frac{\partial T_{\rho}(s_1, s_2)}{ \partial s_1} +    \frac{( k_1 H(s_2) + k_2 s_1) }{2 \Omega}  \frac{\partial^2 T_{\rho}(s_1, s_2)}{ \partial s_1^2 } \nonumber \\
    +& ( k_3 H(s_1) - k_4 s_2) \frac{\partial T_{\rho}(s_1, s_2)}{ \partial s_2} +  \frac{( k_3 H(s_1) + k_4 s_2) }{2 \Omega}  \frac{\partial^2 T_{\rho}(s_1, s_2)}{ \partial s_2^2 }.
\end{align}
\end{widetext}
We impose two reflecting boundaries at zero concentration and an absorbing boundary along the separatrix,
\begin{subequations}
\begin{align}\label{conditions}
    T_{\rho}(s_1, s_2) =& 0  \quad \text{if } s_1=s_2 , \\
    \frac{\partial T_{\rho}(s_1, s_2)}{\partial s_1} =& 0 \quad \text{if } s_1=0 , \\
     \frac{\partial T_{\rho}(s_1, s_2) }{\partial s_2}=& 0 \quad \text{if }  s_2=0. 
\end{align}
\end{subequations}
In Fig. \ref{fig9}, we present the mean escape time $T_1(s_1,s_2)$ together with its standard deviation, $\sigma_T(s_1,s_2)$. The mean escape time increases for high concentrations of $s_1$ and low concentrations of $s_2$ (and symmetrically in the opposite basin), where the system lies deeper within the basin of attraction of the stable stationary state, making escape events less probable. Near the diagonal $s_1=s_2$, the escape time decreases because the system starts close to the separatrix. As in Model I, $\sigma_T(s_1,s_2)$ exhibits a spatial structure and magnitude comparable to $T_1(s_1,s_2)$, consistent with nearly exponential intrinsic escape-time distributions \cite{hanggi1990reaction}.

\begin{figure*}
    \centering
    \includegraphics[width=0.325\linewidth]{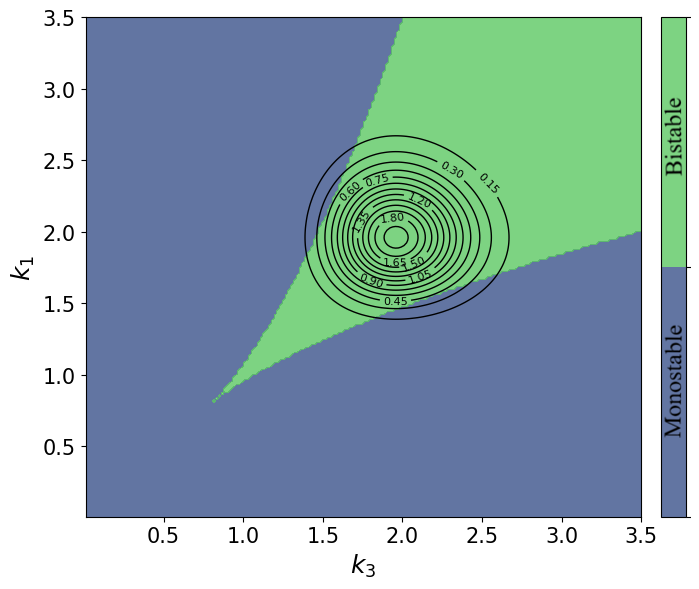}\includegraphics[width=0.325\linewidth]{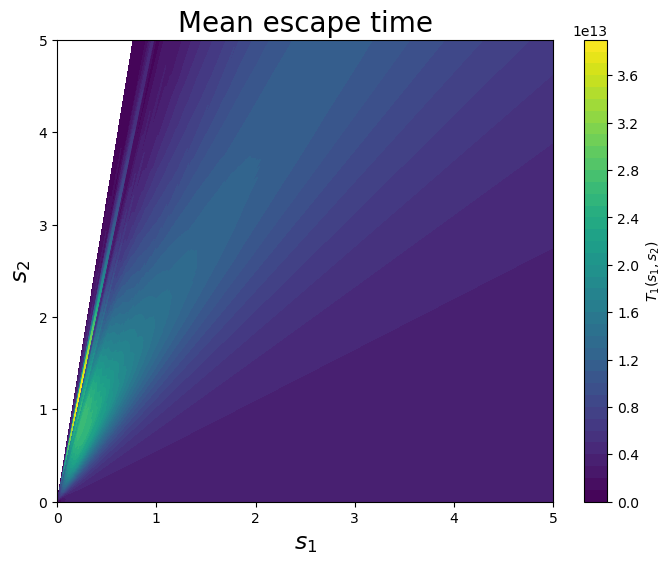}
    \includegraphics[width=0.325\linewidth]{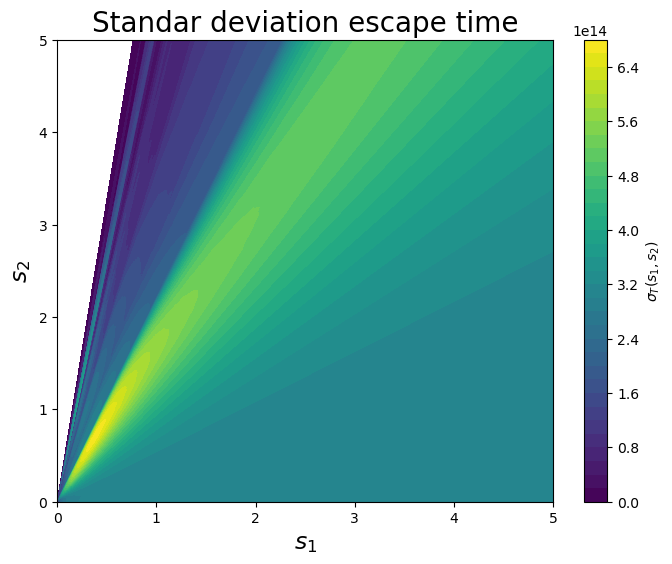}
    \caption{\textbf{Escape time for the toggle switch with intrinsic and quenched extrinsic fluctuations.} In the first panel, we identify the bistable and monostable regions of the $(k_1, k_3)$ parameter space and superimpose the stationary gamma distributions of $k_1$ and $k_3$ defined in Eq.~\eqref{15}. Although part of the probability mass lies in the monostable region, our analysis is conditioned on the bistable subpopulation $\mathcal{B}$. In the last two panels, we present the population mean escape time and its standard deviation. Both quantities increase substantially compared with the purely intrinsic case, with $\sigma_T$ growing much faster than $T_1$, indicating highly heterogeneous switching times across cells. We used $k_{0,1}=k_{0,3}=2$, $k_2=k_4=0.4$, $\Omega=20$, and $a=50$.}
    \label{fig10}
\end{figure*}

\begin{figure*}[h!t]
\includegraphics[width=0.495\textwidth]{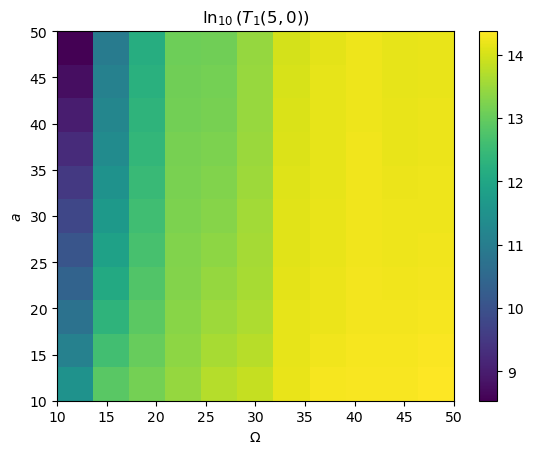}\hfill
\includegraphics[width=0.495\textwidth]{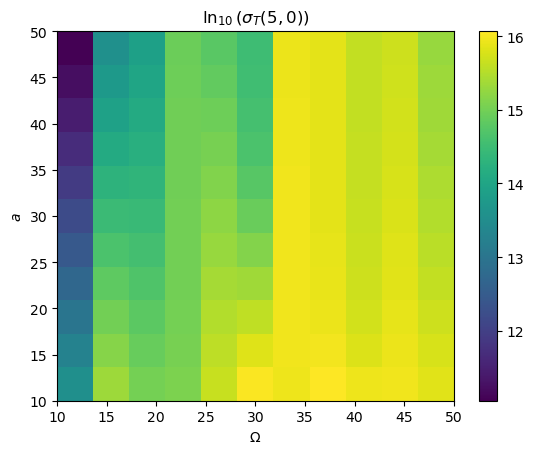}
\caption{\textbf{Escape time for the toggle switch under intrinsic and quenched extrinsic fluctuations.} Mean escape time and its standard deviation (on a $\log_{10}$ scale) starting from $(5,0)$ for different values of $\Omega$ and $a$, when $k_1$ and $k_3$ follow stationary gamma distributions. As the shape parameter $a$ increases, cell-to-cell heterogeneity weakens and the switching-time statistics gradually approach the purely intrinsic values. Parameter values are $k_{0,1}=k_{0,3}=2$ and $k_2=k_4=0.4$.}
\label{fig11}
\end{figure*}

Because our main interest is to calculate the escape time from a stable phenotypic state, the right panel of Fig. \ref{fig9} shows the mean escape time and its standard deviation for trajectories starting from $\bar{s}_2$. Since the system is symmetric, the behavior is identical starting from $\bar{s}_1$. As $\Omega$ increases, both $T_1(\bar{s}_2)$ and $\sigma_T(\bar{s}_2)$ grow exponentially (appearing linear on the $\log_{10}$ scale), reflecting the exponential suppression of intrinsic barrier-crossing events in larger systems, while maintaining $\sigma_T \approx T_1$.

\subsection{Intrinsic and quenched extrinsic fluctuations}

Next, we compute the switching-time statistics in the presence of both intrinsic fluctuation and quenched extrinsic parameter heterogeneity. To model cell-to-cell variability, we assume that the synthesis parameters $k_1$ and $k_3$ are independently drawn from the stationary gamma distribution of a slow Cox--Ingersoll--Ross process \cite{Vastola} (Eq.~\eqref{15}; see Appendix \ref{A}). We restrict extrinsic variability to these two parameters because they are directly associated with transcriptional synthesis rates.

Following the same procedure as in Model I, we solved Eq. \eqref{toggle_switch_time} for each bistable pair $(k_1,k_3) \in \mathcal{B}$, where for asymmetric realizations ($k_1 \neq k_3$) the absorbing separatrix $\delta\Sigma^a(k_1, k_3)$ tilts away from the diagonal $s_1 = s_2$. For simplicity, we assigned identical marginal gamma distributions to $k_1$ and $k_3$, with population means $k_{0,1}=k_{0,3}=2$ matching the purely intrinsic case, and integrated the resulting moments according to Eq. \eqref{11}.

As shown in the left panel of Fig. \ref{fig10}, the \textit{Toggle Switch} exhibits both bistable and monostable regimes in $(k_1, k_3)$ space. Because the gamma distributions extend into the monostable region, a fraction of cells in the population exhibit monostable dynamics rather than bistable switching---a feature related to the coexistence of monostable and bistable cellular subpopulations \cite{Biga, Coomer}. Following Section \ref{section3}, we condition the population average in Eq.~\eqref{11} on the bistable subpopulation $\mathcal{B}$ via $\tilde{p}_{\mathrm{ss}}(k_1, k_3)$.

In Fig. \ref{fig10} (middle and right panels), we show how the mean escape time and its standard deviation are modified by quenched extrinsic heterogeneity. Compared with the purely intrinsic case (Fig. \ref{fig9}), both quantities increase substantially, with the standard deviation exhibiting a particularly pronounced growth ($\sigma_T \gg T_1$). Two mechanisms drive this behavior. First, conditioning on the bistable region $\mathcal{B}$ excludes low-rate monostable realizations while retaining larger $(k_1, k_3)$ values that correspond to deeper basins of attraction and longer escape times. Second, as captured by the law of total variance in Eq.~\eqref{total_var}, cell-to-cell variations in $(k_1, k_3)$ shift both the stationary states and the absorbing separatrix $\delta\Sigma^a(k_1, k_3)$, producing huge dispersion in $T_1(s_1, s_2, k_1, k_3)$ across cells and thus a dominant extrinsic variance term $\mathrm{Var}_{\mathbf{k}}[T_1]$.
 
In Fig. \ref{fig11}, we present the mean escape time and its standard deviation for different values of $\Omega$ and $a$, starting from the point $(5,0)$ near the stable stationary state $\bar{s}_2$. For a fixed $\Omega$, both $T_1$ and $\sigma_T$ decrease as $a$ increases and gradually approach the purely intrinsic values, reflecting the narrowing of the cell-to-cell parameter distribution. Both quantities also increase with $\Omega$ as intrinsic fluctuations become weaker. Across the explored parameter range, $\sigma_T$ remains substantially larger than $T_1$, indicating a highly heterogeneous distribution of phenotypic switching times: while some cells switch rapidly, others remain in their initial phenotype for orders of magnitude longer. Such variability can contribute to phenotypic heterogeneity within genetically identical populations, leading to diverse cellular responses to environmental cues, stress, or therapeutic interventions \cite{geiler2013details, niepel2009non}.

\section{Conclusions} \label{section6}

In this study, we formulated a theoretical framework to compute phenotypic switching-time statistics in bistable GRNs with positive feedback, combining the backward chemical Fokker--Planck equation with population averaging over quenched (slowly varying) extrinsic parameter heterogeneity ($\tau_k \gg T_1$). Because biological cell populations are inherently heterogeneous, accounting for both intrinsic biochemical fluctuation and cell-to-cell parameter variability is essential for obtaining a realistic description of stochastic phenotypic switching.

We applied the proposed framework to two paradigmatic bistable motifs: positive autoregulation and the genetic Toggle Switch. In both systems, quenched extrinsic heterogeneity substantially modified the population-level switching dynamics. In particular, conditioning on the switching-competent bistable subpopulation, extrinsic parameter variability increased the mean escape time and, more importantly, produced a pronounced growth in the standard deviation ($\sigma_T \gg T_1$), yielding much broader switching-time distributions than in the purely intrinsic case.

Our results show that the impact of extrinsic heterogeneity is not limited to adding minor dispersion around the intrinsic mean. Because cell-to-cell variations in kinetic parameters shift the positions of the stable and unstable stationary states as well as the absorbing separatrix, the conditional mean switching time varies exponentially across cells. Through the law of total variance, this parameter-induced variability dominates the total variance and generates highly heterogeneous switching kinetics across genetically identical cells.

From a biological perspective, this heterogeneity has important implications. Wide dispersion in phenotypic switching times implies that genetically identical cells exposed to the same macroenvironment may transition between phenotypes on vastly different timescales. Such variability can contribute to phenotypic diversification, bet-hedging, stress adaptation, and the emergence of rare subpopulations, playing a key role in cellular differentiation, antibiotic persistence and resistance, and drug-tolerant states in cancer and microbial populations.

More broadly, our findings demonstrate that cell-to-cell extrinsic heterogeneity should be treated as a primary determinant of stochastic switching dynamics rather than a small perturbation to intrinsic fluctuation. Natural extensions of this work include incorporating dynamical extrinsic fluctuation with finite correlation time $\tau_k$, accounting for intercellular coupling relevant to collective tissue-level processes, and benchmarking the Fokker--Planck approximation against discrete stochastic simulations in the small-$\Omega$ regime.

\begin{acknowledgments}
Manuel E. Hernández-García acknowledges the financial support of SECIHTI through the program "Becas Nacionales 2023".\\
Natalia López-Paleta acknowledges the financial support of SECIHTI through the program "Becas Nacionales 2021". 
\end{acknowledgments}

\section*{Declarations}
The authors declare no conflicts of interest regarding the publication of this article. 

All data generated or analyzed in this study are included in this published article.

\appendix
\section{Relationship between the Langevin equation and the Forward FPE} \label{A}

Because the dynamics of the kinetic parameters can be formulated either through a Langevin equation or its corresponding forward FPE, we summarize the connection between both descriptions and their stationary distributions \cite{Gar, Vastola}.

Consider a one-dimensional It\^o Langevin equation for a kinetic parameter $k(t)$,
\begin{align}
    \frac{dk}{dt} = f(k) + \sqrt{F(k)}\, \epsilon(t),
\end{align}
where $f(k)$ and $F(k)$ are the drift and diffusion functions, respectively, and $\epsilon(t)$ is Gaussian white noise satisfying $\langle \epsilon(t) \rangle = 0$ and $\langle \epsilon(t)\epsilon(t') \rangle = \delta(t-t')$. The corresponding forward FPE is \cite{Gar}
\begin{align}
    \frac{\partial P(k,t)}{\partial t} = -\frac{\partial}{\partial k}\left[ f(k) P(k,t) \right] + \frac{1}{2} \frac{\partial^2}{\partial k^2}\left[ F(k) P(k,t) \right]. \label{FP1}
\end{align}
We focus on two canonical models with correlation time $\tau_k$: the Ornstein--Uhlenbeck (OU) process and the Cox--Ingersoll--Ross (CIR) process \cite{Vastola}.

First, the OU process is described by the Langevin equation
\begin{align}
    \frac{dk}{dt} = \frac{k_0 - k}{\tau_k} + \sqrt{\frac{k_0^2}{a \tau_k}}\, \epsilon(t), \nonumber
\end{align}
where $k_0$ is the deterministic steady-state value, $\tau_k$ is the relaxation (correlation) time, and $a$ controls the fluctuation intensity. Identifying $f(k) = (k_0 - k)/\tau_k$ and $F(k) = k_0^2/(a \tau_k)$, the corresponding forward FPE is
\begin{align}
    \frac{\partial P(k,t)}{\partial t} = -\frac{1}{\tau_k}\frac{\partial}{\partial k}\left[ (k_0 - k) P(k,t) \right] + \frac{k_0^2}{2 a \tau_k} \frac{\partial^2 P(k,t)}{\partial k^2}. \nonumber
\end{align}
With initial condition $P(k,0) = \delta(k - k_0)$, its time-dependent solution is
\begin{align}
    P(k,t) = \sqrt{\frac{a}{\pi k_0^2 (1 - e^{-2t/\tau_k})}} \exp\!\left[ -\frac{a (k - k_0)^2}{k_0^2 (1 - e^{-2t/\tau_k})} \right], \nonumber
\end{align}
whose stationary solution ($t \to \infty$) is a normal distribution independent of $\tau_k$:
\begin{align}
    P_{\mathrm{ss}}(k) = \sqrt{\frac{a}{\pi k_0^2}} \exp\!\left[ -\frac{a (k - k_0)^2}{k_0^2} \right].
\end{align}

Second, for the CIR process, the It\^o Langevin equation is
\begin{align}
    \frac{dk}{dt} = \frac{k_0 - k}{\tau_k} + \sqrt{\frac{2 k_0 k}{a \tau_k}}\, \epsilon(t),
\end{align}
with $f(k) = (k_0 - k)/\tau_k$ and $F(k) = 2 k_0 k / (a \tau_k)$. Its corresponding forward FPE reads

{\small
\begin{align}
    \frac{\partial P(k,t)}{\partial t} = -\frac{1}{\tau_k}\frac{\partial}{\partial k}\left[ (k_0 - k) P(k,t) \right] + \frac{1}{\tau_k} \frac{\partial^2}{\partial k^2}\left[ \frac{k_0 k}{a} P(k,t) \right]. \nonumber
\end{align}}
For $P(k,0) = \delta(k - k_0)$, the solution is
\begin{align}
    P(k,t) = c_t \left( \frac{k}{k_0 e^{-t/\tau_k}} \right)^{\frac{a-1}{2}} e^{-c_t (k + k_0 e^{-t/\tau_k})} \nonumber \\
     \times I_{a-1}\!\left(2 c_t \sqrt{k k_0 e^{-t/\tau_k}}\right), \nonumber
\end{align}
where $I_{a-1}$ is the modified Bessel function of the first kind of order $a-1$ and $c_t = a / [k_0 (1 - e^{-t/\tau_k})]$. Its stationary distribution ($t \to \infty$) is a gamma distribution, also independent of $\tau_k$:
\begin{align}
    P_{\mathrm{ss}}(k) = \frac{\left( \frac{a}{k_0} \right)^{a}}{\Gamma(a)} k^{a-1} e^{-\frac{a}{k_0} k},
\end{align}
where $\Gamma(a)$ is the gamma function. Because $P_{\mathrm{ss}}(k)$ is independent of $\tau_k$, the quenched extrinsic fluctuation limit ($\tau_k \gg T_1$) naturally preserves the same stationary gamma distribution across the cell population \cite{biondo2023out, keizer2019extending}.


\bibliography{apssamp}

\end{document}